\documentclass[
superscriptaddress,
 reprint,
 showpacs,preprintnumbers,
 nofootinbib,
 amsmath,amssymb,
 aps,
 prd,
 twocolumn
]{revtex4-1}

\usepackage{graphicx}
\usepackage{amsmath}
\usepackage{amssymb}
\usepackage{dcolumn}
\usepackage{bm}
\usepackage{hyperref}
\usepackage[capitalise]{cleveref}
\usepackage[utf8]{inputenc}
\usepackage{color}
\usepackage{booktabs}
\usepackage{aas_macros}
\usepackage[normalem]{ulem}

\newcommand{\be}{\begin{equation}}
\newcommand{\ee}{\end{equation}}
\newcommand{\bea}{\begin{eqnarray}}
\newcommand{\eea}{\end{eqnarray}}
\renewcommand\({\left(}
\renewcommand\){\right)}

\AtBeginDocument{
\heavyrulewidth=.08em
\lightrulewidth=.05em
\cmidrulewidth=.03em
\belowrulesep=.65ex
\belowbottomsep=0pt
\aboverulesep=.4ex
\abovetopsep=0pt
\cmidrulesep=\doublerulesep
\cmidrulekern=.5em
\defaultaddspace=.5em
}

\begin{document}

\title{First Simulations of Axion Minicluster Halos}

\author{Benedikt Eggemeier}
\email{benedikt.eggemeier@phys.uni-goettingen.de}
\affiliation{Institut f\"ur Astrophysik, Georg-August-Universit\"at G\"ottingen, D-37077 G\"ottingen, Germany}
\author{Javier Redondo}
\email{jredondo@unizar.es}
\affiliation{CAPA \& Departamento de F\'{i}sica Te\'{o}rica,  Universidad de Zaragoza, 50009 Zaragoza}
\affiliation{Max-Planck-Institut f\"{u}r Physik,  D-80805 M\"{u}nchen, Germany}
\author{Klaus Dolag}
\affiliation{Max-Planck-Institut f\"ur Astrophysik, D-85741 Garching, Germany}
\affiliation{University Observatory Munich, D-81679 M\"unchen, Germany}
\author{Jens C. Niemeyer}
\affiliation{Institut f\"ur Astrophysik, Georg-August-Universit\"at G\"ottingen, D-37077 G\"ottingen, Germany}
\affiliation{Department of Physics, University of Auckland, Private Bag 92019, Auckland, New Zealand}
\author{Alejandro Vaquero}
\affiliation{Department of Physics and Astronomy, University of Utah, Salt Lake City, Utah, 84112, USA}

\date{\today}

\begin{abstract}
We study the gravitational collapse of axion dark matter fluctuations in the post-inflationary scenario, so-called axion miniclusters, with N-body simulations. 
Largely confirming theoretical expectations, overdensities begin to collapse in the radiation-dominated epoch and form an early distribution of miniclusters with masses up to $10^{-12}\,M_\odot$. After matter-radiation equality, ongoing mergers give rise to a steep power-law distribution of minicluster halo masses. The density profiles of well-resolved halos are NFW-like to good approximation. The fraction of axion dark matter in these bound structures is $\sim 0.75$ at redshift $z=100$. 
\end{abstract}

\maketitle

The QCD axion is a hypothetical particle predicted in the Peccei-Quinn (PQ) mechanism for solving the strong CP problem, and is considered one of the best motivated dark matter (DM) candidates~\cite{PhysRevLett.40.223,PhysRevLett.40.279,PhysRevLett.43.103,shifman,DINE1981199,PRESKILL1983127,PQ1,PQ2}. 
In the so-called postinflation scenario, the axion field takes initial conditions after a phase transition happening after cosmic inflation, and its resulting DM density distribution has large fluctuations on subparsec comoving scales. 
Their gravitational collapse results in the formation of so-called axion miniclusters (MCs) 
with characteristic masses and radii of order $M_\mathrm{mc}\sim 10^{-12}\,M_\odot$ and $R_\mathrm{mc}\sim 10^{12}\,\mathrm{cm}$~\cite{Hogan1988,Kolb:1993zz,Kolb:1994fi,Kolb:1995bu}, a range\footnote{Note that the estimates depend strongly on the cosmological assumptions before big bang nucleosynthesis~\cite{Nelson:2018via,Visinelli:2018wza}.} in which they could be detected in femto-, pico-~\cite{Kolb:1995bu} and microlensing surveys~\cite{fairbairn2018}. 
Moreover, the clumping of DM axions in bound objects has a direct implication in the direct detection at terrestrial experiments~\cite{Tinyakov2016,OHare:2017yze,Knirck:2018knd}  and could have an impact in indirect  detection~\cite{Tkachev:2014dpa,Pshirkov:2016bjr}, see also~\cite{Tkachev:2015usb}. Thus, quantitative predictions for the distribution of axions and the properties of MCs in this scenario are important. 

The evolution of axion DM can roughly be split into three separate stages. The first encompasses the evolution of the axion field from PQ symmetry breaking until after the QCD phase transition when the axion mass has reached its low-temperature value, but well before the onset of gravitational instability. It is governed by the formation and decay of topological defects and nonlinear field dynamics. This early-universe epoch has recently been investigated with special focus on MC formation by means of large lattice simulations \cite{Vaquero_2019,Buschmann:2019icd}. During the second stage, gravity takes over as the dominant force while scalar field gradients can be neglected on the scales of density perturbations, allowing their description with N-body methods for collisionless fluids~\cite{Zurek:2006sy}. Semianalytic tools for structure formation can be employed to predict the properties of minicluster halos (MCHs) such as the minicluster halo mass function (MC-HMF) \cite{fairbairn2018}. Finally, MCHs evolve into large-scale DM halos and become the sites of galaxy formation in the third epoch. Tidal disruption of MCHs and the formation of axion streams are of particular importance during this final stage in order to predict the statistics of axion clumping at the present time \cite{Tinyakov2016,Dokuchaev:2017psd}.

\begin{figure*}
    \centering
    \includegraphics[width=0.7\textwidth]{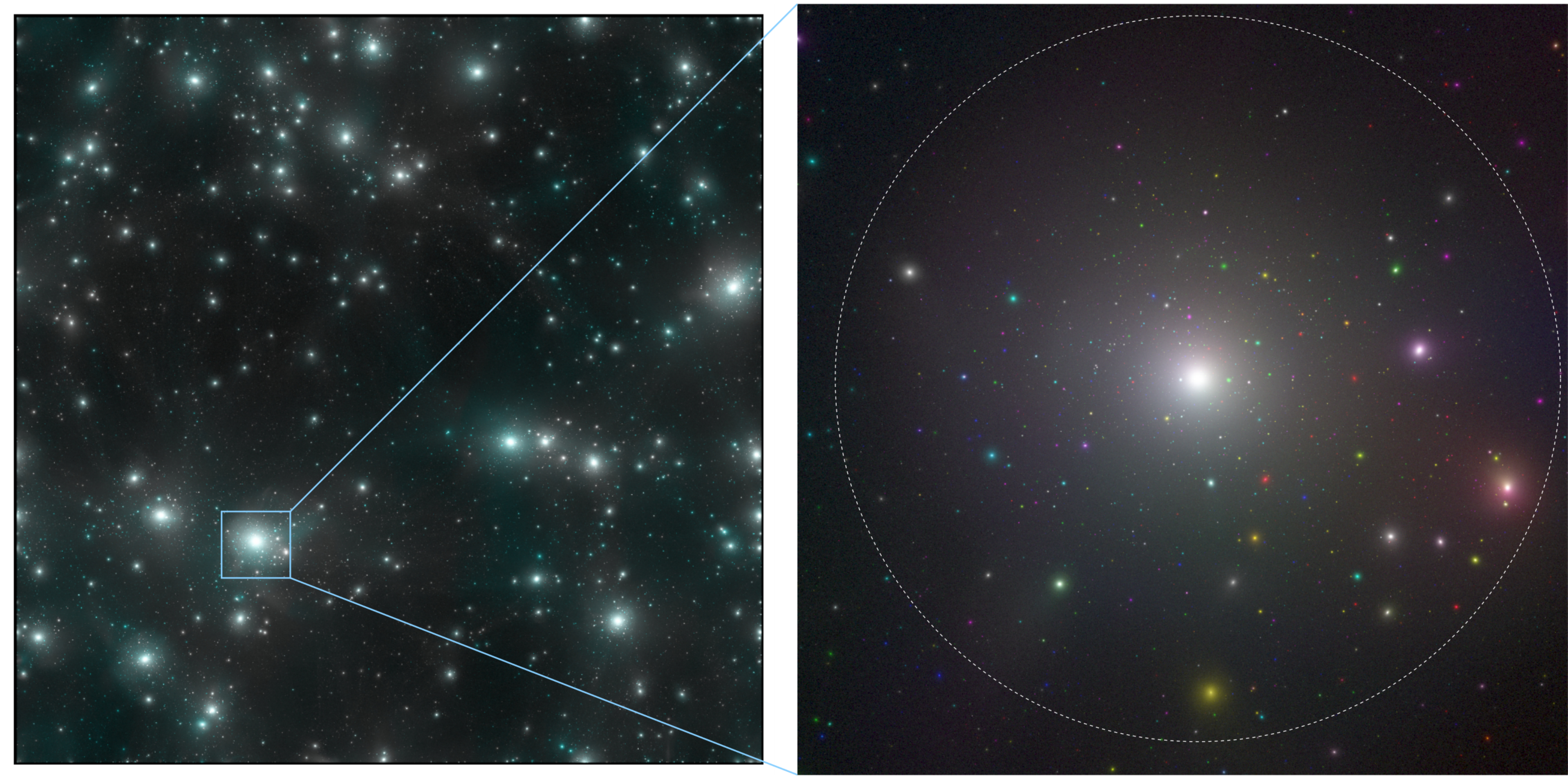}
    \caption{Left: projected axion density of the full simulation box at $z=99$. Right: an enlargement of the largest MCH, where the dashed circle indicates the sphere with density $\rho = 200\,\rho_{m,0}$. The sub-MCs are colored according to their orbital velocity. }
    \label{fig:fullbox}
\end{figure*}

This Letter reports the first results from large N-body simulations addressing the second stage of this process, the formation of axion MCHs by gravitational collapse of primordial axion density perturbations. In particular, we discuss the evolution of the MC-HMF, the fraction of axions bound into MCHs and the MCH density profile. More detailed statistics will be presented in a follow-up publication.



\emph{Simulations of axion density perturbations.--}
We start from initial conditions produced by early-universe simulations using the methods described in Ref.\ \cite{Vaquero_2019}. 
The frozen density distribution resulting from the evolution of the axion field at redshift $z\simeq 10^6$ was converted to $1024^3$ particles in a box with comoving side length $L=0.864\,\mathrm{pc}$  and periodic boundary conditions. The length corresponds to $24L_1$ where $L_1= 2  (1+z(t_1))t_1$ is the comoving coherence length of the axion field at the time $t_1$ when its mass starts to dominate its dynamics ($m_A(t_1)t_1=1/2$, see Appendix~\ref{sec:early-univ}). For simplicity, we assume that axions account for the total amount of DM.

We follow the gravitational evolution of the system with the \textsc{Gadget}-3 code to a final redshift determined by the time when perturbations on the scale of the computational volume become nonlinear (see \cref{sec:early-univ,sec:PSlin,sec:N-body} for details). A visualization of the full simulation box at the final redshift, $z_f = 99$, is shown in \cref{fig:fullbox}. An enlargement of the largest halo reveals its rich substructure.  

MCHs are defined as clusters of gravitationally bound particles in close analogy with DM halos in simulations of structure formation. We identify and characterize them by their virial masses and radii using the \textsc{Subfind} halo finder \cite{Springel2001}. At $z_f=99$, the masses and radii span the ranges $2.5\times 10^{-16}\sim 3.0\times 10^{-9}\,M_\odot$ and $0.4\sim 92.0\,\mathrm{AU}$, respectively.



\emph{Minicluster halo mass function.--}
The MC-HMF is the comoving number density of gravitationally bound MCHs per logarithmic mass interval as a function of MCH mass. It provides a quantitative picture of the dynamics of MCH formation. 
\begin{figure*}
    \centering
    \includegraphics{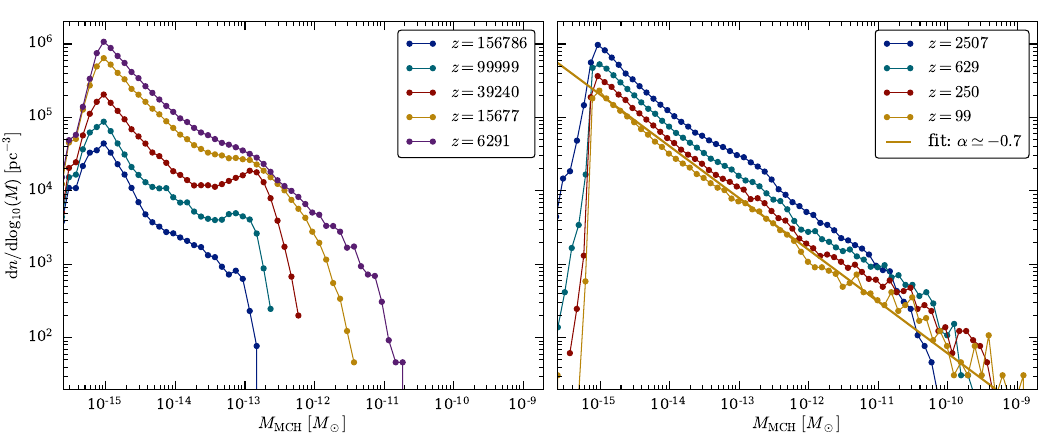}
    \caption{MC-HMF at different redshifts $z$ separated into times before (left) and after matter-radiation equality (right). The slope of the MC-HMF at $z_f=99$ is $\alpha\simeq-0.7$.}
    \label{fig:hmf}
\end{figure*}

The MC-HMF computed from our simulation for different redshifts is shown in \cref{fig:hmf}. 
At early times ($z \gg z_\mathrm{eq}$, left panel), the MC-HMF grows quickly. 
It is dominated at first by halos near the low-mass resolution cutoff $\sim 10^{-15}\,M_\odot$ 
and develops a pronounced peak at $M_\mathrm{mc} \sim 10^{-13}\, M_\odot$ by $z \simeq 4 \times 10^4$. This rapid growth can be understood as the collapse of the density fluctuations that are deeply nonlinear at high-$z$.  
Thus, we can identify the peak as due to the largest nonlinear fluctuations, which should be the ``canonical'' MCs. The abundance of low-mass MCs is the result of the small density seeds found in~\cite{Vaquero_2019} when simulating axions with strings. 
The overall amplitude of the MC-HMF rises until matter-radiation equality, flattening out the peak at $M_\mathrm{mc}$ while extending toward higher masses. 

By the time of equality ($z \simeq z_\mathrm{eq}$), the MC-HMF has developed into a power-law with a slope
of $\alpha \simeq -0.7$ and an exponential cutoff at $\sim 10^{-11}M_\odot$, corresponding to the largest canonical MCs, which typically had only $\mathcal{O}(1)$ initial  overdensities~\cite{Vaquero_2019}. 

During the postequality evolution ($z \ll z_\mathrm{eq}$, right panel in \cref{fig:hmf}) the high-mass cutoff continues to grow at the expense of the total amplitude, which smoothly declines in time. Fitting the MC-MHF to a power-law times a high-mass cutoff still prefers the same overall slope $\alpha \simeq -0.7$. 
However, the fluctuations that collapse after $z_{\rm eq}$ are already small (linear) and the semianalytic Press-Schechter method predicts a MC-HMF $\mathrm{d}n/\mathrm{d}\mathrm{log}\,M \propto M^{-1/2}$~\cite{Enander:2017ogx,fairbairn2018}, which is also compatible with the high-mass data. 
Indeed, a double power-law fit with cutoff provides a better fit to the MC-HMF in this regime. More statistics are needed to quantify it, which we leave for future work. 

\begin{figure}
    \centering
    \includegraphics{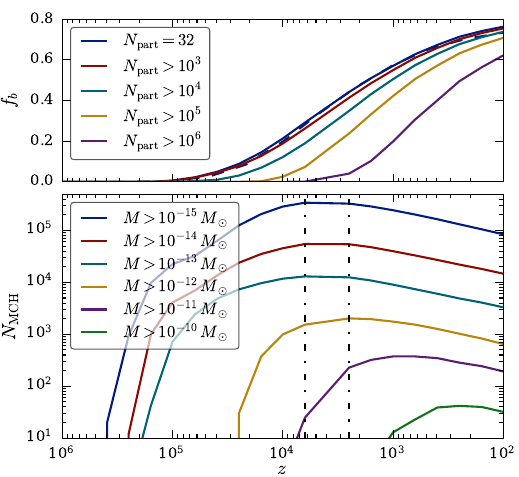}
    \caption{Top: mass fraction $f_b$ of gravitationally bound axions as a function of redshift $z$ considering MCHs with at least $N_\mathrm{part}$ particles as seen in the legend. Convergence of mass resolution for the $N_\mathrm{part}=32$ case is shown by comparing simulations with $1024^3$ and $512^3$ particles (blue dotted line). Bottom: evolution of the total number of MCHs $N_\mathrm{MCH}$ above different mass scales as seen in the legend. The black dotted lines mark the transition from the radiation-dominated to the matter-dominated epoch.}
    \label{fig:mb}
\end{figure}

The late evolution is dominated by mergers with slowly diminishing accretion of unbound axions onto existing MCHs. This is confirmed by the slow saturation of the total fraction of bound axions (upper panel of \cref{fig:mb}) reaching $f_b\sim 0.75$ (taking into account MCHs with at least 32 particles) at $z_f = 99$, and the evolution of the total number of MCHs $N_\mathrm{MCH}$ (lower panel of \cref{fig:mb}). 
Considering only MCHs with at least $10^3$ particles, the evolution of $f_b$ and the final result at $z_f=99$ do not change significantly. At the final redshift $60\%$ of all axions are bound in MCHs with more than $10^6$ particles.
Apart from this, we see that after their formation at $z\simeq 7\times 10^5$  
the number of MCHs grows until $z \simeq z_\mathrm{eq}$. Afterwards, their number is reduced as a result of ongoing mergers. By distinguishing between $N_\mathrm{MCH}$ above certain mass scales we observe at which redshift MCHs with increasing masses emerge. Evidently, MCHs with masses up to $10^{-11}\,M_\odot$ begin to form before matter-radiation equality while higher-mass MCHs arise only for $z< z_\mathrm{eq}$.

\begin{figure}
    \centering
    \includegraphics{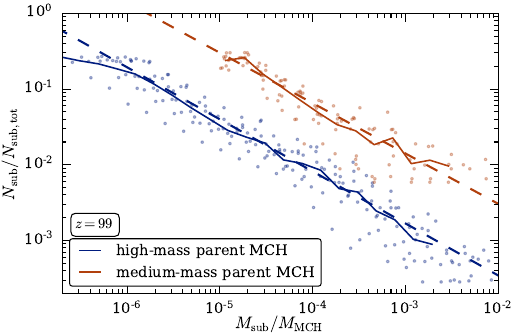}
    \caption{Sub-MC-HMFs of ten high-mass (blue data points) and medium-mass (red data points) MCHs normalized to the virial mass of the parent MCH at redshift $z=99$. The solid lines 
    represent the average of the combined data for the high-mass and low-mass MCHs, respectively. 
    The dotted lines are power-law fits to the data, both consistent with $\alpha\simeq-0.7$.}
    \label{fig:substructure}
\end{figure}

In order to characterize the distribution of sub-MCs within the MCHs, we compare the substructure of ten high-mass MCHs with ten medium-mass MCHs (mass samples are defined in \cref{tab:samples}) in \cref{fig:substructure}. For this, we identified all sub-MCs within the virial radius of each MCH and normalized the sub-MC masses to the virial mass of the corresponding parent MCH. 
\Cref{fig:substructure} shows the relative number of sub-MCs, i.e. the number of sub-MCs divided by the total number $N_\mathrm{sub,tot}$ of sub-MCs contained within the parent MCH. 
For both subsets, the slopes of the averaged sub-MC-HMFs 
are similar to that of the MC-HMF,  $\alpha\simeq -0.7$. The independence of the slopes from the parent MCH mass agrees with previous results for subhalo mass functions in cold dark matter (CDM) simulations~\cite{Lucia2004,Dolag2009}.



\begin{figure}
    \centering
    \includegraphics{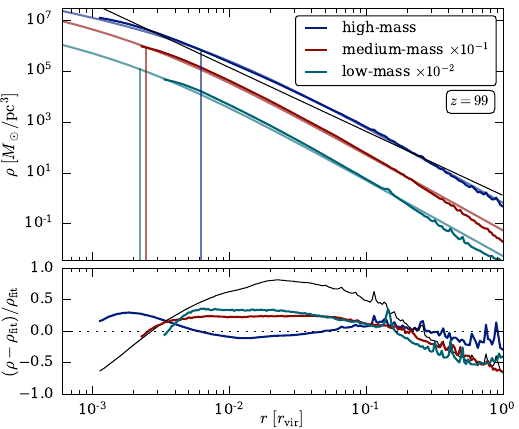}
    \caption{Top: averaged radial density profiles of 20 miniclusters in each mass bin (dark solid lines) truncated at a radial distance of 4 times the numerical softening length. The light solid lines represent NFW fits, where vertical lines mark the corresponding scale radii. The thin black line shows the best-fit power law $\rho\sim r^{-2.52}$. Bottom: deviations from the fit shown in the upper panel.}
    \label{fig:nfw_fitrange}
\end{figure}

\emph{Density profiles.--} 
We study the angular-averaged density profiles $\rho(r)$ of MCHs 
in the last snapshot of our simulation, $z_f = 99$, for which we separated them into three mass samples (cf. \cref{tab:samples}).  
\begin{table}[h]
    \caption{Selected mass samples of MCHs, their respective concentration parameter from an NFW fit and its sensitivity to the radial fit range (details in the text).}
    \centering
    \begin{tabular}{@{}lcccc@{}}
    \toprule
        & $M_\mathrm{MCH}$ [$10^{-11}\,M_\odot$] & $r_\mathrm{vir}$ [AU] &$c$ & sensitivity\\ \midrule
         high-mass & $26-300$ & $40.8 - 92.0$ & $160$ & $3\%$\\
         medium-mass & $3.4-4.6$ &$20.7 - 22.8$& $400$ & $6\%$\\
         low-mass &  $\sim 0.8$ &$\sim12.7$& $450$ &$11\%$\\
    \bottomrule
    \end{tabular}
    \label{tab:samples}
\end{table}
The stacked density profiles of 20 MCHs in each sample, truncated at a radial distance of 4 times the numerical softening length, are plotted in \cref{fig:nfw_fitrange} (upper panel) together with their best-fit Navarro-Frenk-White (NFW) parameterizations given by \cite{NFW} 
\begin{align}
    \rho_\mathrm{NFW}(r) = \frac{\rho_0}{r/r_s(1+r/r_s)^2}\,,
    \label{eq:nfw_profile}
\end{align}
where $\rho_0$ is the characteristic density of the halo and $r_s$ the scale radius.
For comparison, we also show the best-fit power-law for the high-mass MCHs.
As seen in the lower panel of \cref{fig:nfw_fitrange}, 
high-mass MCHs are in good agreement with NFW profiles across the entire radial range, and the scale radius is well resolved. 
The medium-mass and low-mass MCHs, however, are slightly underdense at large radii $r\sim r_{\rm vir}/2$, and the scale radii from the NFW fits are close to or even below the spatial resolution limit. The deviations of the outer density profiles from the NFW fits can be possibly explained by an increased mass accretion as discussed in \cite{Diemer2014}.

The resulting concentration parameter, $c = r_\mathrm{vir}/r_s$, is of the order of several $10^2$ (cf. \cref{tab:samples}) and increases for decreasing MCH masses, in agreement with CDM N-body simulations~\cite{Navarro2004}. In order to examine the stability of the fits, their radial range was reduced by $5\%$, which varies the concentration parameter of the high-mass and medium-mass sample by a few percent. The increased sensitivity for the low-mass sample is related to the fact that the scale radius is only resolved  
for the high-mass and the medium-mass MCHs. 
The MCHs from the low-mass sample together with MCHs of masses down to $10^{-13}\,M_\odot$, which make up $9\%$ of the total number of MCHs 
above the low-mass resolution cutoff, have a density profile consistent with the outer $r^{-3}$-slope of NFW profiles. 
A verification of its convergence to \cref{eq:nfw_profile} would require higher mass resolution and will be addressed in a follow-up publication. 
Nevertheless, we conclude that the density profiles at $z=99$ do not match a $\rho\sim r^{-9/4}$ power-law predicted for spherical accretion from a homogeneous background~\cite{bertschinger1985}. Instead, our results are consistent with high-resolution simulations of ultracompact minihalos producing NFW density profiles for even mild deviations from spherical symmetry \cite{UCMH:gosenca2017}. 

Studying the MCH density profiles at earlier times, we observe that they slowly converge to NFW profiles. The detailed evolution, as well as questions concerning possible differences between MCHs and MCs formed from mergers or monolithic collapse, are left to future work.


\bigskip

\emph{Discussion.--}
We have studied the formation of axion MCs and their clustering into MCHs from  postinflationary symmetry breaking initial conditions. Our results are based on the highest resolution simulations performed to date, both for the initial conditions and their gravitational evolution. The main conclusions are a nearly scale-invariant MC-HMF with slope $\alpha \simeq -0.7$, density profiles that converge toward an NFW shape for $z \ll z_\mathrm{eq}$ at least for sufficiently massive MCHs with concentration parameters of an order of several $10^2$, and a final bound fraction of $f_b \sim 0.75$. 

Of these, the bound fraction is the least robust prediction for axion DM at $z=0$. Improving it will require a better understanding of tidal interactions with stars in the Milky Way. For example, current estimates for tidal disruption by stellar encounters scale with the mean MC density \cite{Tinyakov2016,Dokuchaev:2017psd}, which is an ambiguous concept for MCHs with NFW-like density profiles. We hope that our results provide a starting point for better models, as the problem is probably intractable for full simulations.

More work is also needed to explore the morphology of MCHs, including their mass-dependent substructure and evolution of density profiles as a function of redshift. In particular, it is plausible that features of the ``original'' MCs that clustered into MCHs remain distinguishable even at late times.

Finally, let us consider the predicted population of axion stars in the context of MCHs. Recent studies have shown that the formation of axion stars in the cores of MCHs is a firm prediction \cite{Levkov:2018,Eggemeier:2019}. Following the relation between the mass of the axion star and the host MCH found in simulations \cite{Schive:2014hza,Veltmaat:2018dfz,Eggemeier:2019}, $M_\ast \sim M_\mathrm{MCH}^{1/3}$, we can expect axion stars with masses ranging from $10^{-17}-10^{-15}\,M_\odot$.  
Although the mass ranges of the axion stars and the smallest identified MCHs overlap, we note that our simulations are not capable of resolving them. This is because the value of the de Broglie wavelength $\lambda_\mathrm{dB} = (m v)^{-1}$ that determines the scale of the axion star radius does not exceed the numerical softening length across the entire mass range of the MCHs. 

\section*{Acknowledgements}
We thank Richard Easther, Mateja Gosenca, Shaun Hotchkiss, Doddy Marsh, Bodo Schwabe and Jan Veltmaat for useful discussions and comments. JR acknowledges support from Grants Nos. PGC2018-095328-B-I00 (FEDER/Agencia estatal de investigaci\'on) and FSE-DGA2017-2019-E12/7R (Gobierno de Arag\'on/FEDER) and by a Mercator Fellowship in  Germany‘s Collaborative Research Center (SFB 1258). 
KD acknowledges support by the DFG Cluster of Excellence ORIGINS. JCN acknowledges funding by a Julius von Haast Fellowship Award provided by the New Zealand Ministry of Business, Innovation, and Employment and administered by the Royal Society of New Zealand. AV acknowledges support by the U.S. National Science Foundation under Grant No. PHY14-14614. The simulations where performed at the Leibniz-Rechenzentrum under project ``pr74do''.
BE, JR and AV acknowledge the hospitality of the Munich Institute for Astro- and Particle Physics, 
funded by the Deutsche Forschungsgemeinschaft under Germany's Excellence Strategy EXC-2094--390783311, 
during the 2020 ``Axion Cosmology" workshop where some of the work in this paper was done.  

\appendix

\section{Early Universe simulation}
\label{sec:early-univ}
Numerical  simulations have been widely used to describe the evolution of the axion field from the very early Universe through the period where it becomes non-relativistic and starts behaving as a coherent-state of very cold particles (temperatures $T_1\sim \rm GeV$ in a radiation dominated pre-BBN Universe). 
However, only recently two groups have presented very detailed studies on the evolution and final distribution of DM~\cite{Vaquero_2019,Buschmann:2019icd} including for the first time the effects of cosmic strings and domain walls. 
 Albeit the dynamical range available in the simulation is very far from physical, the current direct results appear largely insensitive to it, and thus on the ensuing small string tension. 
This could very well be a result of the small tensions  themselves~\cite{Gorghetto:2018myk} but at least, an effective model reaching physically relevant high tensions~\cite{Klaer:2017qhr} shows that this parameter does not affect much the DM yield~\cite{Klaer:2017ond}. 
Further work to understand the dynamics of high-tension strings and axion DM is required to clarify these issues, see~\cite{Fleury:2015aca,Fleury:2016xrz,Klaer:2017qhr,Klaer:2017ond,Gorghetto:2018myk,Kawasaki:2018aa,Drew:2019mzc} for recent studies.   

Because of the fast increase of the axion mass with decreasing $T$ the DM distribution at the large scales of interest freezes rapidly below $T_1$. 
The resulting axion DM density distribution has density fluctuations with a standard deviation \mbox{$\delta \rho_A/\bar \rho_A \simeq 0.45$} at distances\footnote{Here $\sigma$ is the width of the Gaussian window function. } $\sigma = L_1$, where $L_1$ is set by the horizon size at $T_1$~\cite{Vaquero_2019},
\begin{align}
\label{L1}
    L_1 = \frac{1}{a_1H_1} = 0.0362\left(\frac{50\,\mu\mathrm{eV}}{m_A}\right)^{0.167}\,\mathrm{pc}\,.
\end{align}

As $\sigma$ grows encompassing more correlation lengths, the white noise fluctuations in the number of $\sim L_1^3$ make the fluctuations decrease as $\propto \sigma^{-3/2}$. 
At small scales, fluctuations grow until they saturate around $\delta \rho/\bar \rho \sim \sqrt{3}$ at the smallest scales, 
see Fig. 21 of~\cite{Vaquero_2019}. 
Note that when our field-simulations end, pseudo-breathers called axitons would continue to evolve to increasingly smaller objects but they are expected to diffuse away to a considerable extent after the axion mass saturates and are thus irrelevant for the $\sim L_1$ scales of interest here~\cite{Vaquero_2019}. 

The size of our simulation will constrain the minimum redshift at which we can trust our gravitational evolution and the maximum ``typical" mass of our MCHs. This is because periodic boundary conditions start to play a significant role when fluctuations of the order of the box size become nonlinear and respond to their ``periodic" copies. In order to be able to reach small redshifts we require large boxes. 
Using a large box also has the advantage to increase the statistics of typical mass MCHs. 
In order to evolve to $z_f \simeq 99$, we produced early-universe axion field simulations with $L=24 L_1$ in $8192^3$ grids using the techniques of~\cite{Vaquero_2019}. Simulating larger boxes compromises the resolution of the string cores or the requirements of sufficient tension to avoid nonphysical destruction of domain-walls by string creation. 

At the end of the simulation the axion field is evolved with the linearised equations (using the WKB approximation) until the redshifts of interest $z\sim 10^6$, similarly to~\cite{Buschmann:2019icd}. This process accounts for the free-streaming of axions until the redshift of interest but is only relevant for the highest-momentum axions. Therefore only the smallest scales of our 8192$^3$ grid are softened.  

The linear growth of gravitational perturbations of a scalar field is generally hampered by the ``quantum" or gradient pressure at length scales smaller than the comoving axion Jeans wavelength,  \begin{eqnarray}
    \lambda_J &=& \frac{2\pi}{\(16\pi G \bar \rho_a(t_0)/(1+z)\)^{1/4}m_a^{1/2}, }\\ 
    &\sim &  \nonumber
    \left(\frac{m_a}{10^{-4} \rm eV}\right)^{-1/2}
    \left(\frac{1+z}{10^4}\frac{0.12}{\Omega_a h^2}\right)^{1/4}\,\mathrm{mpc}, 
\end{eqnarray}
which is smaller than our resolution at all times and only comparable at the initial time $z_i\simeq 10^6$ deep in the radiation domination epoch where gravity is still frozen for our moderate perturbations. 

In order to sample our density field into particles, we first smooth the density into a 1024$^3$ grid of $\sim \mathrm{mpc}$ grid-spacing. 
We then compute the density normalised to the average $n_i=\rho_i/\bar \rho$. The sum of the normalised density is by definition $\sum_i n_i =1024^3$ so we create a number of particles equal to $floor(n_i)$ and distribute them around the grid point $i$ with coordinates displaced by a Gaussian probability distribution with standard deviation equal to half the grid spacing $L/2048$. We decide whether to put a last particle or not by sampling a binomial distribution with probability $n_i-floor(n_i)$. This procedure does not produce exactly a set of $1024^3$ particles -- in this case it felt short by 28783 -- but the sampling is adequate for our purposes. 
We checked that the resulting dimensionless power spectrum $\Delta^2$ coincides with the original grid up to momenta $k\sim 1500$ pc$^{-1}$ (cf. \cref{fig:ps}), above which white-noise from discretization kicks in. Note that at these scales $\Delta^2$ is already decreasing.

The velocities of the particles at $z=10^6$ should reflect the original free-streaming, denoted $v_f$, plus the gravitational acceleration exerted from $z\sim 10^{12}$ until $z=10^6$ during the so-called ``linear evolution" (see below), $v_g$. Both $v_f, v_g$ can be estimated assuming they would be independent,
\begin{eqnarray}
\label{eq:comino}
        v_f &=& \frac{k/a}{m_A} = 3.53 \times 10^{-12} \frac{1+z}{10^6}\left(k L_1\right)\left(\frac{50\mu\rm eV}{m_A}\right)^{0.833}\\
        \langle |{\bf v}_{g}|^2 \rangle &\sim& \left(\frac{3}{2}H\frac{a^2}{a_{\rm eq}}\right)^2\int \frac{dk}{k}\frac{\Delta^2}{k^2} \sim \left(\frac{3}{2}H\frac{a^2}{a_{\rm eq}}\right)^2 L_1^2\,,\\
        \label{eq:gravino}
    |v_{g}| &\sim& 3\times 10^{-10}
    \left(\frac{50\mu\rm eV}{m_A}\right)^{0.167}\frac{z_{\rm eq}}{4000}\,.  
\end{eqnarray}
We have already argued that the density modes available in our grid are below the Jean's length at $z\sim 10^6$. Indeed our estimate of the free-streaming velocity $v_f$ is also smaller than $v_g$ for the modes available after smoothing because $(k L_1)_{\rm max}=2\pi\times 512/24\sim 134.0$. Essentially, both approaches are the same and thus offer the same conclusion: free-streaming is irrelevant at $z\sim 10^6$ in our simulation. 
The gravitational component has been estimated from the variance of density fluctuations assuming a constant density field at $z\gtrsim 10^6$. This is reasonable because the gravitational potential is suppressed by a higher power of $k$ in the denominator and thus it is mostly sensitive to modes on $1/L_1$-scales, which have been completely  frozen already at $z\gtrsim 10^{8}$ according to Eq.~\eqref{eq:comino} and Eq.~\eqref{eq:gravino}. 
In the linear and radiation dominated regime, the solution to Euler's equation, $\dot {\bf v} + H{\bf v} + \nabla \Phi/a=0$ (here $a$ denotes the scale factor $a=1/(1+z)$) for the peculiar velocity field of the DM fluid with a constant density contrast $\delta(x)$ solves to 
\begin{equation}
{\bf v}({\bf x}) = ({\bf v}_i({\bf x})-{\bf v}_g({\bf x}))\frac{a_i}{a} + {\bf v}_g({\bf x})\,, 
\end{equation}
where ${\bf v}_g = -1.5 Ha^2/a_{\rm eq}\nabla(\nabla^{2}\delta(x))$ is the asymptotic velocity gained through gravity.  By $z\sim 10^6$ the free-streaming part should be negligible and we should have ${\bf v}\sim  {\bf v}_g$. 
For simplicity, we use ${\bf v}_i =0$ at $z_i = 10^6$ and let the gravity dynamics of the N-body code to accelerate $\bf v$ to its asymptotic value $\bf v$, which happens only after a few e-folds. This can only affect the smallest and densest structures, those experiencing significant delays in their collapse with respect to the case where the initial velocities were set to ${\bf v}_g$. In the best possible case, this would simply imply that our smallest structures are a bit less dense than they should. In the worst case, the smallest size and small mass halos collapse at the same time as slightly larger structures, giving their mass to them and producing effectively a small-mass cut-off, like the one we find at $M\sim 10^{-15}M_\odot$. 
Since we are mostly interested on more massive MCHs, we can safely ignore these effects. 
However, it is clear that treating correctly the initial velocity field, also by including the effects of free-streaming in the gravitational potential, will be crucial to study the MC-HMF below the $10^{-15}M_\odot$ limit, the densest MCHs or the MC-HMF with a precision better than $1\%$ (roughly the fraction of $M\lesssim 10^{-13}M_\odot$ MCHs that collapse before $z\sim 10^5$ in our simulations, see Fig.~\ref{fig:hmf}). 
We will do so in a further publication.

\section{Power spectrum and Linear evolution}
\label{sec:PSlin}

\begin{figure*}
    \centering
    \includegraphics{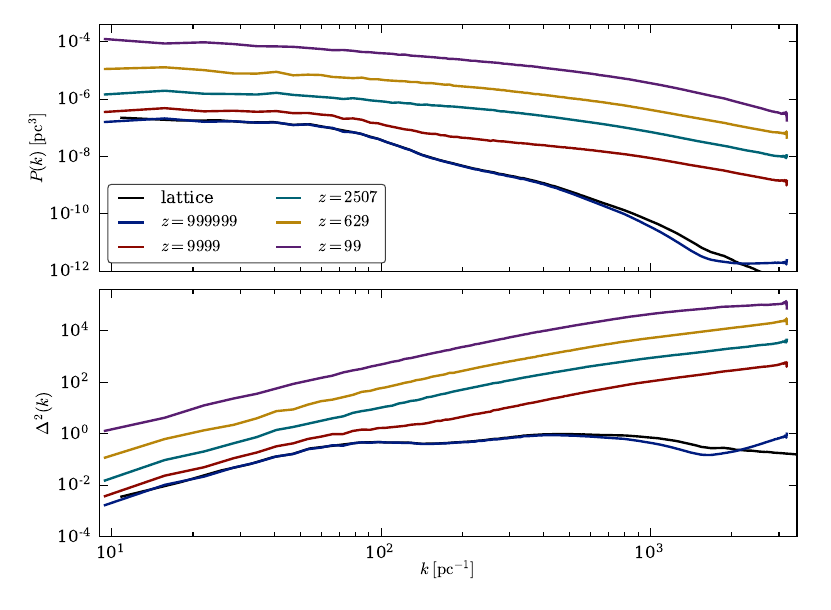}
    \caption{Power spectrum both in physical (upper panel) and dimensionless (lower panel) units for different redshifts. The black line shows the power spectrum directly calculated from the original density field for comparison.}
    \label{fig:ps}
\end{figure*}

The initial dimensionless power spectrum of density fluctuations is a white-noise power law $\Delta^2\propto k^3$ at large scales $k\ll k_1=1/L_1$ saturating around $1/L_1$ and slowly decreasing, cf. \cref{fig:ps} (down). The large dynamical range of $1024^3$ particles allows our simulation to start probing the decrease of $\Delta^2$ at small scales. 
At large scales the variance of the fluctuations within a Gaussian window-function follows the power law $\langle \delta_\sigma^2 \rangle =0.019(5)(L_1/\sigma)^3$ as a function of the Gaussian width $\sigma$. 
Fourier modes of the density field corresponding to large scales are small and evolve linearly through our simulation, as 
\begin{equation}
\delta_k \propto 1 + \frac{3}{2}\frac{1+z_{\rm eq}}{1+z},
\end{equation}    
where we have assumed $\partial_t\delta_k = 0$ at our initial time (which sets  the well-known $\log a$ growth of matter perturbations to zero during radiation domination). 
The largest scales in our simulation $\sigma \sim L/2$ would then become non-linear at a redshift where $\langle \delta_\sigma^2(z_\sigma) \rangle \sim 0.1$ which gives $z_\sigma\sim 60$ for $\sigma=12 L_1$. When these modes become non-linear, the box starts effectively reacting to the gravitational potential of the periodic "copies" of our box  outside it and our simulations can no longer be trusted. 

Power spectra of mass density fluctuations in physical and in dimensionless units are shown in \cref{fig:ps} for different redshifts. 
Comparing the power spectra at the beginning of the simulations and at $z=9999$, we observe
enhanced growth of high-$k$ modes.
As can be seen from the dimensionless power spectrum, the scales of the length of the box start to become nonlinear at $z=99$. Hence, our simulations are reliable until this point.

\section{Comparison to adiabatic perturbations}
The size of the axion isocurvature density fluctuations $\Delta^2(k)=0.03(1) (k L_1)^3$ becomes comparable to the $\sim$scale-invariant adiabatic density fluctuations assumed from inflation, $\Delta^2\sim 2.2\times 10^{-9} T^2 (k/0.05 {\rm Mpc}^{-1})^{0.969-1}$, at a wavenumber $k\sim 3.7\,\mathrm{pc}^{-1}$ (we used a value of the transfer function $T\sim 240$ at the wavenumbers of interest). We would need to simulate boxes $\sim 2$ times larger to start being sensitive to those scales. 
The isocurvature fluctuations that we simulate here correspond to sizes and densities of minicluster seeds. The adiabatic fluctuations originate from the temperature fluctuations, which shift ``locally" the time $t_1$. At large scales, they correspond to a very small  overall up or downwards shift in the axion content of each minicluster seed in our simulation.

\section{N-body simulations and Halo finder}
\label{sec:N-body}

We used the OpenMP/MPI optimized developer version of \textsc{Gadget}-3 which is a successor of \textsc{Gadget}-2~\cite{Springel2005}. The simulations were performed with $1024^3$ particles having a mass of $2.454\times10^{-17}M_\odot$. The numerical softening length, which sets the limit of the spatial resolution, was adjusted to be  $1\,\mathrm{AU}/h$ in comoving units, 
slightly below the Power criterion~\cite{Power:2002sw} and even the revisited lower value in~\cite{Zhang:2018nqh} by a factor of $\sim 4$. 

We chose a comoving box side length of $L=0.864$ corresponding to $24 L_1$ with an axion mass of $m_A=50\,\mu$eV in \cref{L1} and evolved the simulations from $z_i \simeq 10^6$ to $z_f = 99$. The background radiation terms (photons and 3 massless neutrino species, although masses within current cosmological bounds will not make a difference at these scales) were explicitly taken into account for the calculation of the Hubble parameter, 
\begin{align}
    H^2(z) = H_0^2\left(\Omega_{m,0}(1+z)^3 
    + \Omega_{r,0}{(1+z)^4} + \Omega_{\Lambda,0}\right)\,.
\end{align}
We used the standard $\Lambda$CDM parameters $\Omega_{m,0} = 0.3, \Omega_{r,0}=8.486\times 10^{-5}$,  $\Omega_{\Lambda,0}=0.7$ and $H_0 = 100 h\,\mathrm{km}\,\mathrm{s}^{-1}\,\mathrm{Mpc}^{-1}$ with $h=0.7$. We do not include baryons. At our very small scales they are tightly coupled to photons and their density fluctuations would be irrelevant. 

MCHs and sub-MCs were identified by deploying the \textsc{Subfind} algorithm~\cite{Springel2001,Dolag2009}. \textsc{Subfind} starts with a halo list identified through the Friends-of-Friends algorithm, applying a linking length
of $l=0.16$ and considering halos with at least 32 particles. In order to estimate the local density at each particle belonging to an identified halo, it is adopted an adaptive kernel estimation based on all particles using 50 neighbors. Starting from isolated density peaks, additional particles are added in sequence of decreasing density to build the sub-halo candidates.  For this, saddle points in the global density field are exploited to disjoint the sub-halo candidates. All of them then undergo an iterative unbinding procedure with a tree-based calculation of the potential, where the Hubble flow is taken into account. Finally, only sub-halos with at least 20 bound particles are considered. 
To calculate the properties of the halos their center is set to the position of the lowest potential. Virial quantities of the halos are then computed as spherical averages using again all particles. Specifically, we used the virial parameter 
\begin{align}
    \Delta_c = (18\pi^2 + 82x - 39x^2)\,,
\end{align}
where $x = \Omega_m(z) - 1$ and 
\begin{align}
    \Omega_m(z) = \frac{\Omega_{m,0} (1+z)^3}{\Omega_{m,0} (1+z)^3 + \Omega_{r,0}(1+z)^4 + \Omega_{\Lambda,0}}\,.
\end{align}

In our range of interest $z\in (10^6-10^2)$,  we find 
$\Delta_c = \Delta_{\rm vir} \sim (50,180)$. 

The virial radius is defined as the radius for which the average density of the MCH matches the virial parameter times the critical density, 
\begin{align}
    \frac{\int_0^{r_{\rm vir}} 4\pi r^2 \rho(r)\,\mathrm{d}r }{4\pi r_{\rm vir}^3/3} = \Delta_{\rm vir} \rho_c.  
\end{align}
where $\rho_c = 3 H^2/(8\pi G)$ with $H$ the Hubble expansion rate. 
The virial mass is the mass contained within that radius, i.e. $M_{\rm vir}= 4\pi/3 \Delta_\mathrm{vir}\rho_c r_\mathrm{vir}^3$.

\bibliography{axion_mc}

\end{document}